# Crossover from Room-temperature Double-channel Ballistic Transport to Single-channel Ballistic Transport in Zigzag Graphene Nanoribbons

Zhao-Dong Chu and Lin He*

Department of Physics, Beijing Normal University, Beijing, 100875, People's Republic of China

**Abstract**

Very recently, it was demonstrated explicitly that a zigzag graphene nanoribbon (GNR) exhibits a crossover of conductance from $G_0$ to $G_0/2$ with increasing the length ($G_0 = 2e^2/h$ is the quantum of conductance) even at room-temperature [Baringhaus, et al. *Nature* 506, 349 (2014)]. Such a result is puzzling as none of previous theories seem to match the experimental observations. Here, we propose a model to explain the crossover from double-channel to single-channel ballistic transport in zigzag GNR. The $sp^3$ distortion of carbon atoms at the GNR edges induces a large spin-orbit coupling on the edges atoms, which enhances spin-flip scattering of edge states of the zigzag GNR. With sufficient spin-flip scattering, the wave-function of edge states becomes a superposition of the spin-up and spin-down components. Then the coupling of the two edges becomes important. This removes the edge degree of freedom in the zigzag GNR and results in the evolution of the conductance from $G_0$ to $G_0/2$ with increasing the length.

**Significance statement**

A very recent experiment reported the observation of the crossover from double-channel ballistic transport to single-channel ballistic transport in zigzag graphene nanoribbon (GNR) with increasing the length. This result has attracted much attention, however, none of previous theories seem to match the experimental observations. Here, we propose a simple model that could explain quite well the puzzling experimental result. According to our analysis, the decrease of conductance from $G_0$ to $G_0/2$ with increasing the probe spacing direct arises from the joint effect of the reduction of the spin degree of freedom and the edge-edge coupling in the GNRs. Our result indicates that the spin degree of freedom plays an important role in the transport properties of graphene systems.

Ballistic transport of electrons has fascinated researchers across various disciplines for several decades[1-7]. Looking for systems where ballistic transport can be observed at room temperature boosts rapid developments of several leading topics in condensed matter physics in the past few years[8-10]. Graphene nanoribbon (GNR) with perfect crystallography is believed to be one of the most promising candidates as a perfect conductor where electrons can travel long distances without dissipation. However, all experiments in lithographically patterned exfoliated GNRs demonstrated that transport is dominated by diffusive mode rather than the ballistic mode[11,12]. This stimulates many groups to develop different methods in the fabrication of high-quality graphene nanoribbons[13,14].

Until very recently, a great breakthrough in the growth of GNRs with crystallographically perfect edges is achieved and room-temperature ballistic transport with a conductance of $G_0$ in the GNRs is realized by Baringhaus and colleagues for the probe spacing $L \leq 160$ nm (Here $G_0 = 2e^2/h$ is the quantum of conductance and the factor 2 comes from the spin degree of freedom)[15]. These zigzag GNRs are grown on the slope of a terraced silicon carbide surface and the transport properties of the GNRs were measured *in situ* with variable probe spacing. Figure 1 summarizes the room-temperature conductance $G$ versus the probe spacing $L$ of the zigzag GNRs in a range of energies around charge neutrality[15]. For $L \leq 160$ nm, a conductance of $G_0$ is observed. With increasing the probe spacing to $L \sim 1$ μm, the conductance decreases gradually to $G_0/2$, then it keeps a robust value of $G_0/2$ for 1 μm ≤ $L$ ≤ 16 μm. The most surprising result of their experiments is the observation of the single quantum mechanical channel for transport in these high-quality zigzag GNRs with 1 μm ≤ $L$ ≤ 16 μm. This result is amazing since that at least an edge-degenerate channel, giving rise to a conductance of $G_0$, should be involved in the transport of the GNRs in a range of energies around charge neutrality point[16,17]. The observed conductance of $G_0/2$, as shown in Fig. 1, is obviously beyond the description of any previous theories[18].

In this contribution, we propose how the single quantum mechanical channel can readily arise in the zigzag GNR. Even under ultraclean high vacuum conditions, there are many adatoms (impurities), such as hydrogen, that hybridize directly with carbon atoms at edges of GNRs. This hybridization-induced $sp^3$ distortion of the carbon atoms leads to a strong enhancement of spin-orbit coupling[19], which consequently enhances spin-flip scattering of edge states of the zigzag GNRs. In a long zigzag GNR, the wave-function of the edge states becomes a superposition state of the spin-up and spin-down components because of sufficient spin-flip scattering. Then the edge-edge coupling becomes important, which removes the edge degree of freedom and leads to the crossover of the conductance from $G_0$ to $G_0/2$.

**Results**

**Hubbard model of zigzag GNR.** At the charge neutrality level of graphene monolayer, there are four identical channels, two from the spin and two from the valley (K and K'), for ballistic transport[20]. In zigzag GNRs, the valley degree of freedom is removed around the charge neutrality point and only spin-polarized channels located at the two edges presenting a conductance of $G_0$ are expected to be observed[16,17]. To explicitly illustrate this, we show electronic structures of a zigzag graphene nanoribbon in Fig. 2. The electronic structures can be described quite well by a tight-binding model Hamiltonian[21-24]

$$H = -t \sum_{\sigma=\uparrow,\downarrow} \sum_{<i,j>} [c^{\dagger}_{i,\sigma} c_{j,\sigma} + H.c] + U \sum_i n_{i,\uparrow} n_{i,\downarrow} . \qquad (1)$$

Here, the first term is the nearest neighbor hopping term on the honeycomb lattice. The operator $c^{\dagger}_{i,\sigma}$ ($c_{j,\sigma}$) creates (annihilates) an electron with spin $\sigma$ at site $i$ ($j$), $t$ = 2.7 eV represents the nearest-neighbor hopping amplitude of the π electrons, and <$i$, $j$> are nearest neighbors on a honeycomb lattice. The second (Hubbard) term represents electron-electron interaction with the strength of on-site Coulomb repulsion $U$ = 6 eV and $n_{i,\sigma} = c^{\dagger}_{i,\sigma} c_{i,\sigma}$ the spin-resolved electron density at atom $i$. According to the Hamiltonian (1), the zigzag GNRs feature a gap with spin-polarized edge states at each edge and the spin polarizations are in opposite directions in the two edges of a GNR[21], as shown in Fig. 2. Additionally, the states of opposite spin orientation in the zigzag GNRs are degenerate in all bands.

**Conductance of a zigzag GNR described by Hubbard model.** In a transport measurement, an external voltage $V$ creates a difference of chemical potential between two contacts $u^- - u^+ = eV$ and generates a net current[25,26]

$$I = I^- - I^+ = \frac{e}{L}\sum_k vf^- - \frac{e}{L}\sum_k vf^+ . \qquad (2)$$

Here $I^-$ and $I^+$ are the currents of left-moving and right-moving electrons respectively, and $f^-$ and $f^+$ are the corresponding Fermi-Dirac distributions describing the electron density of state. With considering

$$\sum_k \to 2 \times \frac{L}{2\pi}\int dk , \qquad (3)$$

the net current can be written as $I = 2 \times \frac{e}{h}(u^- - u^+)$[25,26]. The factor 2 in Eq. (3) comes from the spin degree of freedom. In the zigzag GNRs described by Hamiltonian (1), the factor 2 also refers to the two spin-polarized edges. Consequently we obtain the conductance around the charge neutrality as $G = \frac{I}{V} = \frac{2e^2}{h}$, which could explain quite well the result of the zigzag GNRs with $L \leq 160$ nm, as shown in Fig. 1. An in-plane electric field applied across a zigzag GNR lifts the spin degeneracy of edge states and the GNR could be forced into a half-metallic state[21]. Then only one spin state (either spin-up or spin-down state) at the two edges contributes to the transport around the charge neutrality point. In such a case, the factor 2, referring to the two edge channels in the zigzag GNR, still plays its role in Eq. (3). Consequently, the conductance is still $G_0$. Obviously, the above analysis cannot explain the result of the zigzag GNRs with $L > 160$ nm, not to mention the crossover of the conductance between $G_0$ and $G_0/2$ in the GNRs as a function of the probe spacing, as shown in Fig. 1. The experimental result in Ref. [15] indicates that the probe spacing plays a vital role in the emergence of the single-channel ballistic transport in the GNRs.

**Spin-orbit coupling of edge atoms in zigzag GNR.** To fully understand the puzzling experimental result in Ref. [15], we propose that effect of hybridization-induced $sp^3$ distortion of edge atoms should be taken into account in the transport properties of the zigzag GNRs. The existence of substantial amounts of adatoms (impurities) is inevitable even under ultraclean high vacuum conditions. These impurities hybridize directly with carbon atoms at edges of GNRs and induce a distortion of the graphene lattice from $sp^2$ to $sp^3$ [19,27,28]. Such a distortion of edge atoms results in a large enhancement of the spin-orbit coupling on the edges[3,4,19,29]

$$H_{edge} = \sum_{<i,j>}^{1,N,n} \lambda s^z_{\uparrow,\downarrow} c_i^\dagger c_j + H.c \quad, \tag{4}$$

where $N$ is the number of zigzag chains in the GNR, $s^z_{\downarrow,\uparrow}$ is a Pauli matrix representing the electron's spin, $\lambda = 10$ meV is the strength of spin-orbit coupling, and $n$ is the number of atoms with $sp^3$ orbital in the 1$^{st}$ and $N^{th}$ zigzag chains, *i.e.*, in the two edges of the GNR. Then the total Hamiltonian of the system can be written as

$$H' = H + H_{edge} \quad. \tag{5}$$

The second term of Hamiltonian (5) enhances spin-flip scattering in zigzag GNRs and is expected to affect the transport properties of the edge states dramatically. As an example, if there is an atom at site $i'$ with the $sp^3$ orbital in the 1$^{st}$ zigzag chain, the spin-flip scattering induced by this atom will force the edge states to be a superposition of the spin-up and spin-down components because

$$H_{edge}|\Psi_1\rangle = \sum_i a_i \langle c_{i',\uparrow}|\varphi_{i,\uparrow}\rangle \cdot |c_{i',\uparrow}\rangle + \sum_i \lambda' a_i \langle c_{i',\uparrow}|\varphi_{i,\uparrow}\rangle \cdot |c_{i'+1,\downarrow}\rangle \quad \text{(here}$$

$|\Psi_1\rangle = |\Psi_\uparrow\rangle = \sum_i a_i |\varphi_{i,\uparrow}\rangle$ is the initial ground state of the 1$^{st}$ zigzag chain).

**Edge state in a zigzag GNR with spin-orbit coupling.** For generally, the wave functions of the edge states localized in the 1$^{st}$ and $N^{th}$ zigzag chains can be written as a superposition of the spin-up and spin-down components:

$$|\Psi_1\rangle = a|\Psi_\uparrow\rangle + b|\Psi_\downarrow\rangle,$$

$$|\Psi_N\rangle = a|\Psi_\downarrow\rangle + b|\Psi_\uparrow\rangle \quad . \quad (6)$$

Here $a$ and $b$ are positive normalized parameters. The wave functions with $a$ and $b$ in opposite signs are not the eigenstates of Hamiltonian (5) with a positive spin-orbit coupling. To describe the effect of $sp^3$ distortion of edge atoms, we define $p$ as the ratio of edge atoms with $sp^3$ orbital with respect to all of the edge atoms in GNR. For $p = 0$, we have $a^2 = 1$ and $b^2 = 0$ ($|\Psi_1\rangle = |\Psi_\uparrow\rangle$ and $|\Psi_N\rangle = |\Psi_\downarrow\rangle$). Then Eq. (1) describes quite well the ground state of zigzag GNR, as shown in Fig. 2. When $p \neq 0$, the wave functions of edge states depend on the probe spacing and we can obtain the coefficients of the wave functions by numerically solving the Hamiltonian (5) in real space.

Figure 3(a) shows the coefficients, $a^2$ and $b^2$, as a function of the probe spacing, i.e., $a^2(L)$ and $b^2(L)$, for different $p$. Obviously, the values of $a^2$ and $b^2$ depend on both the probe spacing and the value of $p$. Here we take the result of $p = 0.8$ as an example to illustrate the effect of Hamiltonian (4) on the transport properties of zigzag GNRs. When $L$ is of the order 100 nm or shorter, we still have $a^2 \approx 1$ and $b^2 \approx 0$, i.e., $|\Psi_1\rangle \approx |\Psi_\uparrow\rangle$ and $|\Psi_N\rangle \approx |\Psi_\downarrow\rangle$. It indicates that the effect of spin-flip scattering is negligible in such a length region even with $p = 0.8$. With increasing $L$ from about 100 nm to about 1 μm, $a^2$ ($b^2$) decreases (increases) gradually towards 0.5. For the case that $L > 1$ μm, we have $a^2 \approx b^2 \approx 0.5$, which means that the wave functions of edge states at the two edges become identical and can be written as $|\Psi_1\rangle \approx |\Psi_N\rangle \approx \frac{\sqrt{2}}{2}|\Psi_\uparrow\rangle + \frac{\sqrt{2}}{2}|\Psi_\downarrow\rangle$, as shown in Fig. 3(b). Then there is no spin degree of freedom in the zigzag GNR.

**Edge-edge coupling in a zigzag GNR.** Along with the variation of wave functions of the edge states, the coupling strength between the two edges also changes as a function of the probe spacing. The metallic probes, with their size of the tip much larger than the width of the GNRs, are expected to connect the two edges of the GNRs and therefore act as a medium for the edge-edge coupling during the transport measurements. An edge-state-only model[22] with the form

$$H_{edge-edge} = t' \sum_{<i,j>} c^\dagger_{1,i,\sigma} c_{N,j,\sigma} + H.c \qquad (7)$$

can capture the essential features of this effect. Here $t' = \sum_{<i,j>} <\Psi_1 \cdot I \cdot \Psi_N>$ describes the magnitude of edge-edge coupling and I is the identity matrix. Obvious, the edge-edge coupling is zero for the zigzag GNRs without $sp^3$ orbital in their edge atoms because that the two edge states are orthogonal to each other. However, for the superposition states, $|\Psi_1\rangle = a|\Psi_\uparrow\rangle + b|\Psi_\downarrow\rangle$ and $|\Psi_N\rangle = a|\Psi_\downarrow\rangle + b|\Psi_\uparrow\rangle$, the magnitude of edge-edge coupling depends on the coefficients of the wave functions and will reach its maximum when $|\Psi_1\rangle \approx |\Psi_N\rangle \approx \frac{\sqrt{2}}{2}|\Psi_\uparrow\rangle + \frac{\sqrt{2}}{2}|\Psi_\downarrow\rangle$. Figure 4(a) shows the relationship between $t'$ and the probe spacing of a zigzag GNR with $p = 0.8$. For small probe spacing, i.e., $L < 100$ nm, the edge-edge coupling is rather weak and it is expected that we could obtain the conductance $G_0$ around the charge neutrality. Whereas, in the strong coupling region, i.e., $L > 1$ μm, the factor 2 is removed from Eq. (3) and only a conductance of $G_0/2$ is expected to be observed in the zigzag GNRs around the charge neutrality. Obviously, our model not only explains the single-channel ballistic transport reported in Ref. [15], but also provides a natural and reasonable explanation in understanding the crossover of the conductance between $G_0$ and $G_0/2$ in the GNRs. The joint effect of the reduction of the spin degree of freedom and the edge-edge coupling in the GNRs leads to the decrease of conductance from $G_0$ to $G_0/2$ with increasing the probe spacing, as shown in Fig. 1. According to our calculation, the curve of $a^2(L)$ with $p = 0.7$, as shown in Fig. 1, matches the experimental result quite well.

**Spin-spin correlation in a zigzag GNR.** To further confirm our analysis, we calculate the spin-spin correlation $\chi$ of the edge states for the zigzag GNRs described by Hamiltonian (5). The spin-spin correlation is defined as[30-33]

$$\chi = \langle \Psi | F | \Psi \rangle, \tag{8}$$

where $|\Psi\rangle = \sum_{i,\uparrow,\downarrow}^{n'} |\varphi_{i,\uparrow,\downarrow}\rangle$ is the wave function of the ground edge state, $n'$ is the total number of atoms in one edge (the 1$^{st}$ or the $N^{th}$ zigzag chain), and $F$ is the spin-spin correlation function between two lattice sites $i$ and $j$ following the form (here we calculate that of the $N^{th}$ zigzag chain as an example)[32,33]

$$\begin{aligned} F_{i,j} &= \langle \vec{S}_{i,N} \cdot \vec{S}_{j,N} \rangle + \langle \sigma^z_{i,N} \cdot \sigma^z_{j,N} \rangle \\ &= \langle \vec{S}_{i,N} \cdot \vec{S}_{j,N} \rangle + \langle (n_{i,\uparrow} - n_{i,\downarrow})(n_{j,\uparrow} - n_{j,\downarrow}) \rangle. \end{aligned} \tag{9}$$

Here $\sigma^z_{i,N} = (n_{i,\uparrow} - n_{i,\downarrow})$ denotes the local magnetic moments in the $N^{th}$ zigzag chain, $n_{i,\uparrow}$ ($n_{i,\downarrow}$) is the number of spin-up (spin-down) electrons at site $i$ and $\vec{S}_i = \frac{\hbar}{2} c_i^+ \vec{s}^z_{\uparrow,\downarrow} c_i$. In Fig. 4(b), the solid curve shows the spin-spin correlation as a function of the probe spacing in the $N^{th}$ chain of a zigzag GNR with $p = 0.8$. The value of $\chi$ decreases dramatically with increasing $L$ because of the spin-flip scattering, and it approaches zero when $L > 1$ μm. The spin-spin correlation between the two edges is also shown in Fig. 4(b) (the dashed curve). Its value increases substantially with increasing $L$ and approaches 1 when $L > 1$ μm. Both the spin-spin correlation of each edge and the spin-spin correlation between the two edges show that the spin-flip scattering completely removes the spin degree of freedom of the GNR when the probe spacing is larger than 1 μm. This agrees quite well with the calculated result in Fig. 3(a).

**Discussion**

Here, we should point out that all of our analyses are based on the GNRs with crystallographically perfect edges. A few imperfections of the edges could lead to dissipations of the edge states, which may account for the further reduction of the conductance in the zigzag GNRs with probe spacing larger than 16 μm (Fig. 1).

In summary, we propose a simple model that could explain quite well the puzzling result observed very recently in the zigzag GNR. According to our analysis, the decrease of conductance from $G_0$ to $G_0/2$ with increasing the probe spacing, as observed experimentally in Ref. [15], direct arises from the joint effect of the reduction of the spin degree of freedom and the edge-edge coupling in the GNRs. The hybridization-induced $sp^3$ distortion of the carbon atoms enhances spin-flip scattering of edge states of the zigzag GNRs, which consequently removes the spin degree of freedom in the GNRs with sufficient long probe spacing. Then the wave function of the edge states in these GNRs becomes a superposition of the spin-up and spin-down components. In the near future, more theoretical studies should be carried out to calculate transport properties of the superposition wave-function in zigzag GNRs. On the one hand, this could help us to quantitative understand the transport properties of the zigzag GNRs with probe spacing larger than 16 μm (as shown in Fig. 1); on the other hand, it may provide unprecedented opportunities to uncover novel electronic behaviors in zigzag GNRs.

*Email: helin@bnu.edu.cn

**Acknowledgments**

We thank Prof. Yugui Yao and Prof. Hua Jiang for helpful discussions. We are grateful to National Key Basic Research Program of China (Grant No. 2014CB920903, No. 2013CBA01603), National Science Foundation (Grant No. 11422430, No. 11374035, No. 11004010), the program for New Century Excellent Talents in University of the Ministry of Education of China (Grant No. NCET-13-0054), and Beijing Higher Education Young Elite Teacher Project (Grant No. YETP0238).

**Author contributions**

Z.D.C. performed the theoretical calculations. L.H. conceived and provided advice on the analysis and theoretical calculation. L.H. and Z.D.C. wrote the paper.

**Competing financial interests:** The authors declare no competing financial interests.

**Figure Legends:**

**FIG. 1. Room-temperature conductance $G$ versus the probe spacing $L$ in zigzag GNRs (the solid circles).** The experimental data are reproduced from Ref. [15]. The zigzag GNRs are 40-nanometer-wide and have crystallographically perfect edges. There is a spin-degenerate channel exhibiting a conductance $G = G_0$ for $L \leq 160$ nm. With increasing the probe spacing from 160 nm to 1 μm, the conductance decreases gradually from $G_0$ to $G_0/2$, then, the conductance keeps a robust value of $G_0/2$ for 1 μm ≤ $L$ ≤ 16 μm. The solid curve is the valve of $a^2$ (right-Y axis) calculated according to our model (Eq. (5)), which is explained in main text.

**FIG. 2. Band structure of a zigzag GNR.** (a) The spatial distribution of the charge difference between spin-up and spin-down for the ground states of an ideal zigzag GNR. The red and green balls represent the charge density of spin-down and spin-up electrons, respectively, and the radius of the circle reflects the magnitude of the charge density. (b) The band structure of a zigzag GNR obtained from Eq. (1). It is reproduced from Refs. [21,23].

**FIG. 3. Spatial distribution of wave functions of edge states in zigzag GNR**. (a) The relationship between coefficients, $a^2$ and $b^2$, of edge wave functions and the probe spacing of zigzag GNR for different $p$. (b) The schematic spatial distribution of wave functions of edge states in zigzag GNR with $p \neq 0$. The red and green balls represent the charge density of spin-down and spin-up electrons, respectively, the purple balls represent the superposition state of the spin-up and spin-down components. The radius of the circle reflects the magnitude of the charge density.

**FIG. 4. The edge-edge coupling and spin-spin correlation in zigzag GNR.** (a) The relationship between the magnitude of the edge-edge coupling (normalized) and the probe spacing in a zigzag GNR with $p = 0.8$. (b) The solid curve refers to the relationship between the spin-spin correlation and the probe spacing in a zigzag GNR with $p = 0.8$. The dashed curve represents the spin-spin correlation between the two edges of the same zigzag GNR.

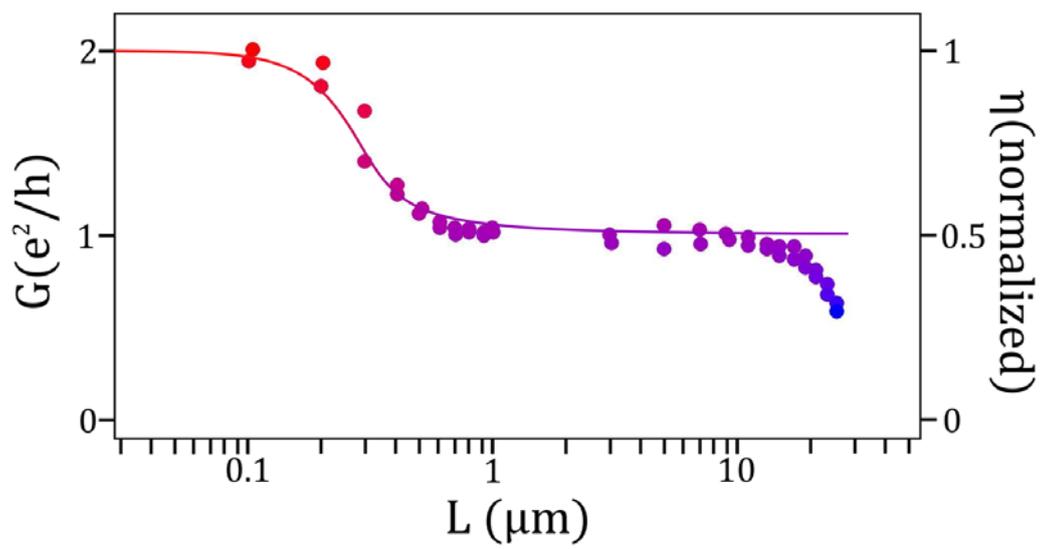

Figure 1

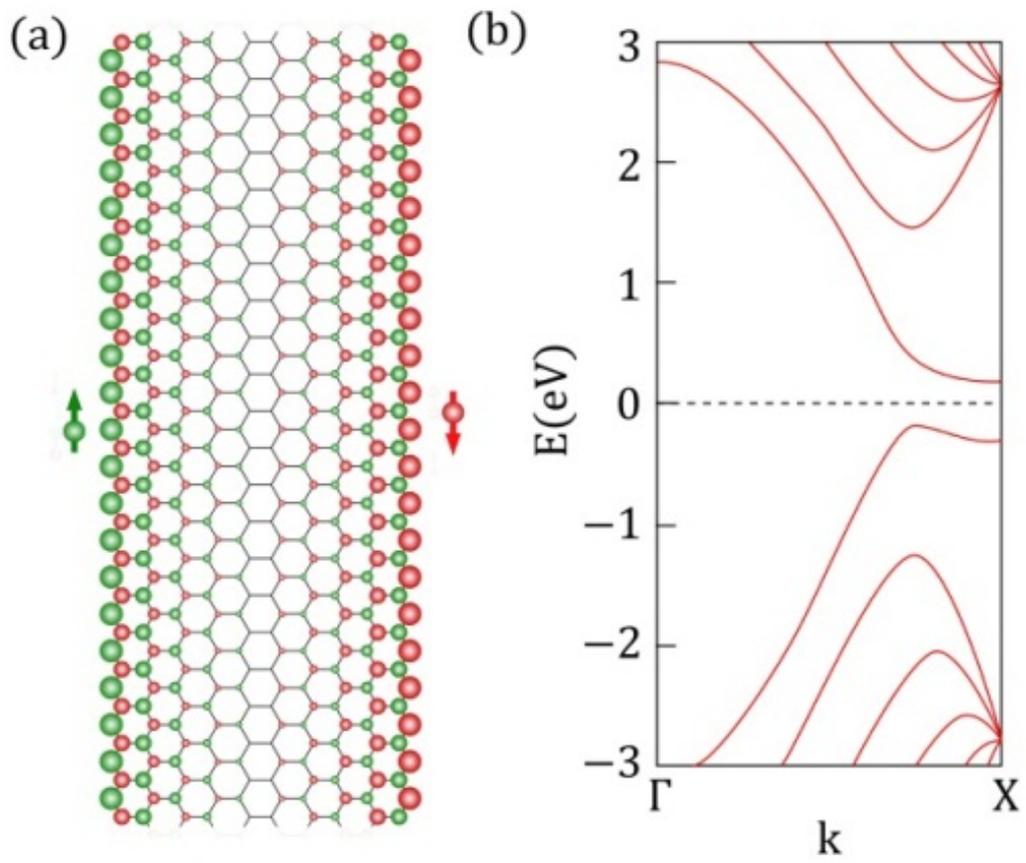

Figure 2

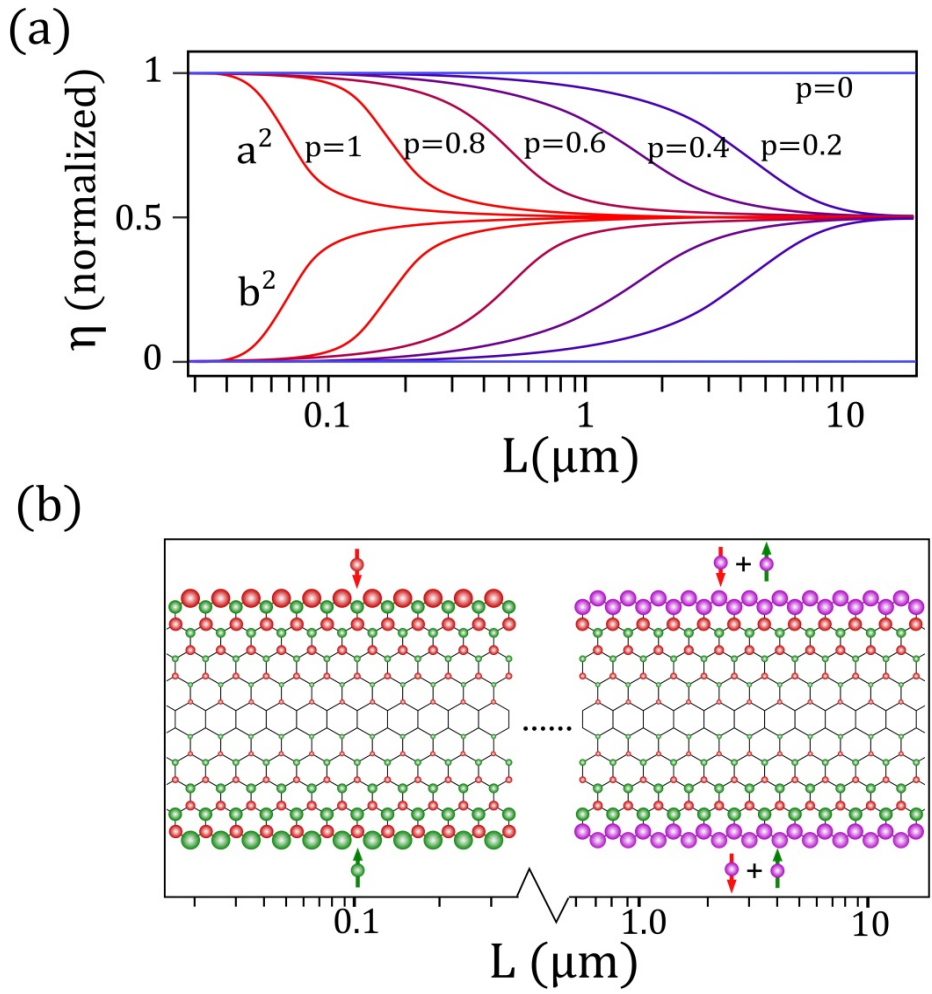

Figure 3

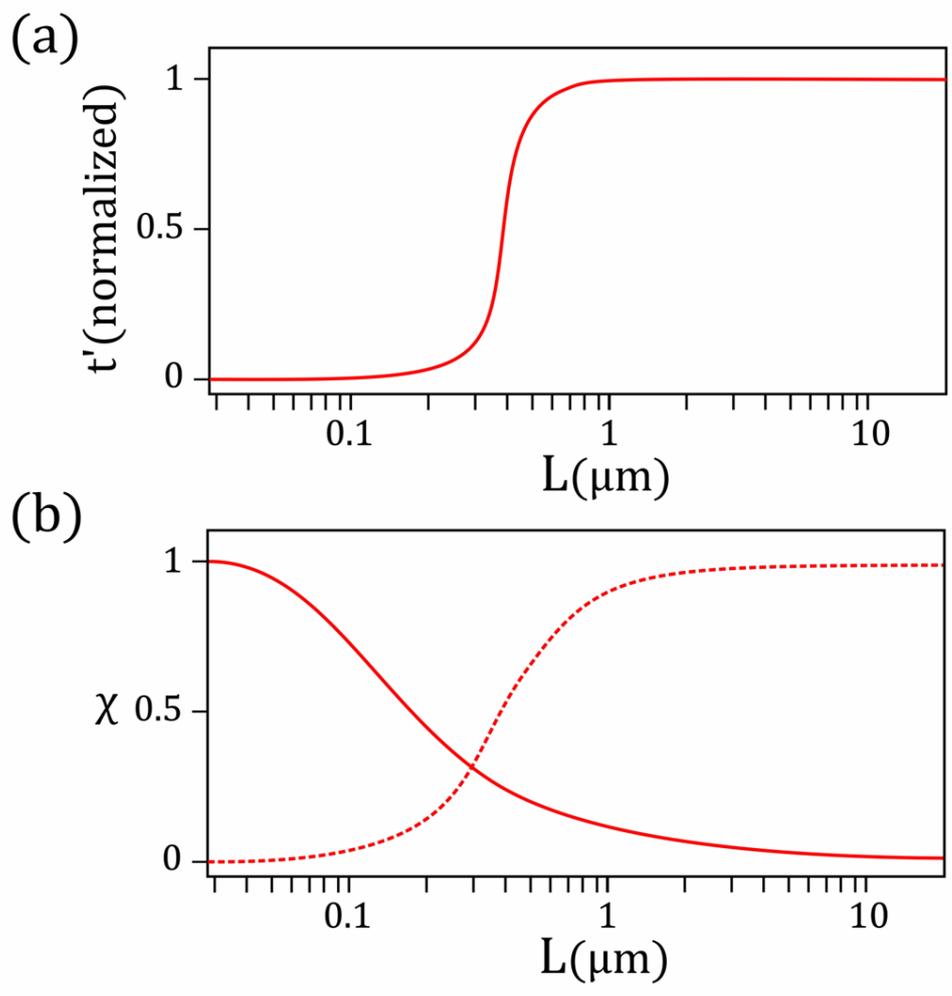

Figure 4